\documentclass[journal]{vgtc}                     

\usepackage{enumitem}
\usepackage{array}
\usepackage{multirow}
\usepackage{makecell}
\usepackage{adjustbox}
\usepackage{tabu}                      
\usepackage{booktabs}                  
\usepackage{lipsum}                    
\usepackage{mwe}                       
\usepackage{comment}
\usepackage{soul}
\usepackage[dvipsnames]{xcolor}
\usepackage{mathptmx}                  
\usepackage{xspace}
\usepackage{amssymb}
\usepackage{amsmath}
\usepackage{mathtools}
\usepackage[skins]{tcolorbox}

\onlineid{XXXX}
\vgtccategory{Research}
\vgtcpapertype{application/design study}

\newcommand{\techname}{VisAutocomplete}

\newcommand{\contributionbox}[1]{%
    \begin{tcolorbox}[
      colback=gray!5,
      colframe=gray!40,
      boxrule=0.4pt,
      arc=2pt,
      left=6pt, right=6pt, top=6pt, bottom=6pt
    ]
  #1
  \end{tcolorbox}
}

\title{Visualization Autocomplete: Visualization Authoring\\ via Stepwise Design Recommendations}

\author{%
    \authororcid{Hyeon Jeon}{0000-0002-9659-2922},  
    \authororcid{Sungbok Shin}{0000-0001-6777-8843}, and
    \authororcid{Niklas Elmqvist}{0000-0001-5805-5301}
}

\authorfooter{
  \item Hyeon Jeon is with Seoul National University, Seoul, South Korea.
  E-mail: hj@hcil.snu.ac.kr.
  \item Sungbok Shin, the corresponding author, is with Sogang University, Seoul, South Korea. 
  E-mail: sbshin90@sogang.ac.kr.
  \item Niklas Elmqvist is with Aarhus University, Aarhus, Denmark. 
  E-mail: elm@cs.au.dk.
}

\abstract{%
    When domain experts create charts, the bottleneck is rarely the data, but knowing the optimal next step in chart design. 
    The visualization design space is vast, and while domain experts can recognize a good design when they see it, it is often challenging to determine the exact path to get there. 
    To address this, we present \textsc{\techname{}}, a system inspired by text autocompletion that reconceptualizes visualization design as a sequential process, recommending concrete next steps at each stage of the authoring process based on common practices.
    Users can intervene at any step, or delegate multiple steps to the system and select one from the design recommendations.
    To support responsive interaction, we distill the translation logic of a large language model (LLM) into a single function that receives the current chart state and recommended transition as input and returns the updated chart specification as output.
    We evaluate the system against a LLM vibecoding, Microsoft Excel, and TaskVis, an automated chart recommendation engine, on chart quality and approachability.
    Our results show that \techname{} outperforms all baselines in the articulacy of complex chart authoring, while remaining on par with LLM in approachability.
}


\teaser{
  \centering
  \includegraphics[width=\linewidth]{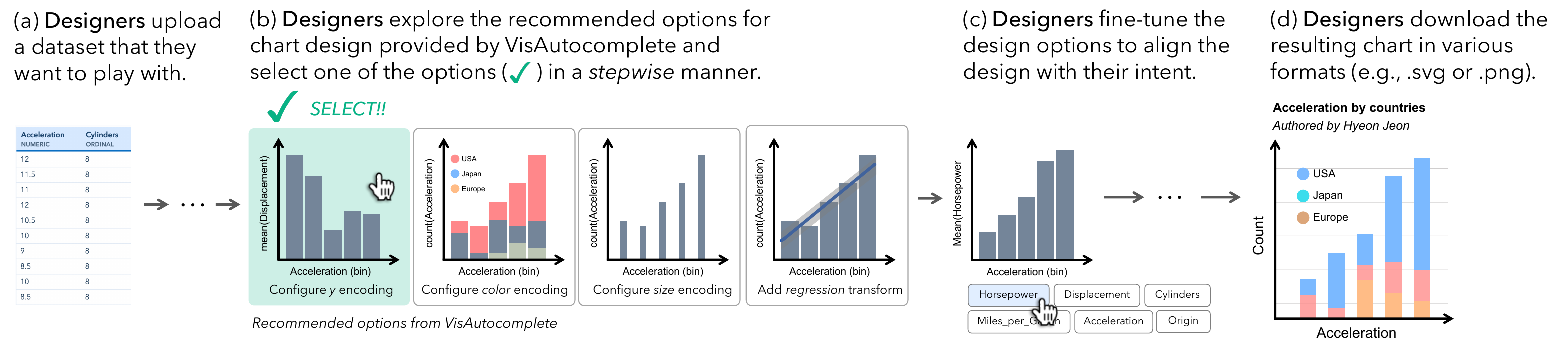}
  \caption{%
        \textbf{Stepwise visualization design with \techname{}.}
        \techname{} is a novel visualization authoring system that supports stepwise design autocompletion using recommendations.
        The workflow starts with authors uploading a dataset (a).
        For each step, the system recommends an ordered list of recommendations to guide the visualization design, where designers can select one among them (b).
        Some recommendations have user-controllable parameters, such as the number of bins, the color scale, or the chart title (c) for minor adjustments. 
        Once the visualization is complete, it can be exported as either a specification or an image file (d).
        }
  \label{fig:teaser}
}

\graphicspath{{figs/}{figures/}{pictures/}{images/}{./}} 

\renewcommand{\paragraph}[1]{\vspace{3pt}\noindent\textbf{#1.}}
\newcommand{\paragraphit}[1]{\vspace{3pt}\noindent\textit{#1.}}

\usepackage{silence}
\newcommand{\rev}[1]{{{#1}}}

\begin{document}

\firstsection{Introduction}

\maketitle

\label{sec:01-introduction}

Effective data visualization is essential for communicating insights quickly and accurately in many disciplines, ranging from science to finance, medicine to mechanical engineering, office work to journalistic inquiry.
However, domain expertise does not necessarily translate to fluency in creating visualizations~\cite{grammel10novicevis}. 
All these decisions, such as choosing the best chart type, variable encoding, color scale, axis layout, or data legend, can easily become overwhelming in the face of a massive array of design possibilities.
While automated visualization design tools exist~\cite{moritz19tvcg, shen21eurovis, wang23acl}, and a subset of these tools have sought a balance between agency and automation~\cite{wongsuphasawat17voyager2}, they tend to require specialized knowledge and notation, and yield resulting visualizations with little opportunity for design tweaking~\cite{lin20dziban, wang21falx, tian25chartgpt} or learning~\cite{gabriel17infovistools} from the process for next time.
Interactive chart design tools~\cite{zong21lyra2, satyanarayan14lyra}, on the other hand, place a considerable burden on users and consequently do not tend to integrate best practice recommendations in their workflows.

To address this gap, we propose \textit{stepwise visualization autocompletion}, an approach that reconceptualizes visualization authoring as a step-by-step process (\autoref{fig:teaser}, \ref{fig:navigation-process}). 
Inspired by text autocompletion that interactively suggests words and phrases as users type, our approach guides users through consecutive visualization design decisions. 
We model each decision as a valid visualization specification. 
At each step, the system recommends a set of possible next states from the current state.
These are derived from design patterns observed in large-scale visualization corpora.

Visualization autocompletion achieves two key qualities simultaneously: \textbf{approachability} and \textbf{articulacy}. 
Automated tools prioritize approachability for domain experts by generating complete visualizations in a single step. 
However, this coarse granularity constrains articulacy, offering little support for delicate expression.
In reality, experts may not always have a clear vision of the final output, and visualization design is inherently an iterative process of exploring possibilities and progressively refining choices. 
Visualization autocompletion achieves both simultaneously by keeping the process approachable by presenting an ordered list of candidate states, and pursuing articulacy by integrating data-driven recommendations with user choice.

We validate visualization autocompletion through a web-based system called \textsc{\techname{}}\footnote{\url{https://vis.autocomplete.studio/}}.
\techname{} guides users through chart creation as a structured, step-by-step authoring process, progressively selecting each component of the visualization specification through recommendations, including data encodings, transformations, and visual styling. 
At each stage, candidate changes are visually previewed and ranked by common practices, which users may accept outright, refine further, or delegate across multiple steps at once. 
Recommendations are generated in real-time \rev{without requiring on-the-fly LLM inference.}
\rev{We implement this by constructing a translation function that takes the current chart specification and a recommended transition as input and returns the updated chart specification. 
This function is authored by the LLM, which we prompt to leverage its code-generation capabilities.} 
At any point, users retain the ability to manually override individual parameters, ensuring that automation enables rather than constrains. 

However, it remains an open question whether such a finer-grained, iterative approach can meaningfully improve articulacy while remaining approachable for domain experts without a background in data visualization.
To evaluate \techname{} across both articulacy and approachability, we compare visualizations produced with the system against those created with popularly used visualization authoring tools in practice: Microsoft Excel and vibecoding using ChatGPT.
For an additional point of comparison, we include TaskVis~\cite{shen21eurovis}, a chart recommendation engine inspired by Draco~\cite{moritz19tvcg}, as a representative of a fully automated baseline. 
We find that \techname{} outperforms Microsoft Excel and automated solutions in terms of chart quality. 
While its performance is comparable to that of the LLMs on creating simple charts, \techname{} demonstrates superior quality when constructing more complex charts.
In terms of approachability, \techname{} is either more approachable than, or at least on par with, the baselines.


    
    
    

\section{Background}
\label{sec:02-background}

Below, we present the background for our research.

\subsection{Visualization Design Process}
\label{subsec:iterative-design-process}

Prior work has sought to characterize the visualization design process~\cite{munzner09nested, sedlmair12designstudymethod, brehmer13tasktypology}, and to understand how designers work in practice~\cite{grammel10novicevis}. 
In particular, researchers have focused on domain experts in professional settings, who have little to no background in data visualization~\cite{Wong2018}. 
A significant proportion of visualization designers fall into this category, and this makes it essential to understand how they engage with the design process in order to better support them~\cite{dvs2024}. 
In practice, designers rarely begin with a clear vision of how their visualization should look~\cite{parsons22designprac}, and the design process is further shaped by unexpected challenges and constraints~\cite{sedlmair12designstudymethod, grammel10novicevis}. 
Consequently, arriving at an effective chart may require significant and iterative exploration of the design space until the designer is satisfied with the result~\cite{satyanarayan20critical}.

Sch\"{o}n's notion of design as a \textit{reflective practice} offers a key insight into how this iterative process actually unfolds: each decision is not planned in advance, but emerges from a \textit{``situated conversation with the current design situation''}~\cite{schon2017reflective, parsons22designprac}---that is, a designer's next move is always a local, grounded response to what they see in front of them at that moment. 
This suggests that while the overall design process may appear nonlinear and unpredictable, it may fundamentally unfold as a sequence of locally situated decisions, each informed by the current state of the design, much like a greedy traversal of the design space, one step at a time. 
This, in turn, suggests the need for a new model that supports the navigation of this iterative process toward a visualization that meets the designer's communicative goals.

\contributionbox{\textbf{Contribution.} Inspired by Schön's notion of reflective practice~\cite{schon2017reflective}, we model visualization design as a stepwise process.}

\subsection{Augmenting Visualization Design}
\label{subsec:chart-authoring-tools}

Several directions have been suggested to help analysts in effective visualization design. 
One line of research automates the design process by recommending charts based on the specifications of the dataset~\cite{wongsuphasawat16tvcg, hu19vizml}.
While this approach reduces the designer's burden, it limits the exploration of the broader design space.
A related line of work incorporates designer intent into the recommendation process. 
\rev{However, these tools are based on programming and specification, which still present a barrier for domain experts without programming knowledge~\cite{moritz19tvcg, lin20dziban}.
A third line of research enhances the expressivity of visualizations via interaction.
These tools originally aim to help novice designers by not having to do programming, but instead through direct user manipulation~\cite{satyanarayan14lyra, zong21lyra2, ren19charticulator, DBLP:conf/chi/LiuTWDDGKS18, DBLP:journals/tvcg/RenHY14}.}
However, they can overwhelm users with complex interfaces, as they are difficult to learn~\cite{satyanarayan20critical}.

Motivated by this, L'Yi et al.~\cite{lyi25blace} propose blended interfaces that link familiar and unfamiliar authoring interfaces to improve learnability.
A different line of tools \rev{assists} designers throughout the design process itself, such as by linting~\cite{hopkins20visualint} or detecting mirages~\cite{mcnutt20mirages}, and by automating design feedback~\cite{shin23scannerdeeply, shin23perceptualpat, shin25visualizationary} to detect issues in chart design.
\rev{Another line of work supports authoring through iterative interaction with generated outputs, whether by comparing design variants~\cite{vdelzen13smallmultiple} or offloading data transformation to an AI agent~\cite{wang24dataformulator}.}

\contributionbox{\textbf{Contribution.}
Rather than recommending a final visualization or exposing users to a complex authoring interface, \rev{we reconceptualize visualization design as a sequential process consisting of primitive actions, and realize it for chart authoring.
At each step, we recommend the most common next design choices, constructing the specification one decision at a time, keeping the interaction simple and learnable.}
}

\subsection{Agency and Automation}
\label{subsec:agency-hcai}

The role of agency in automated systems has been a longstanding topic. 
While early work argues that automation could handle repetitive tasks efficiently~\cite{maes96intelsoft}, researchers later raised concerns about overreliance on it. 
Sheridan delineates the boundaries of appropriate automation use~\cite{sheridan1992telerobotics}, 
Horvitz argues for a synergistic integration of automation and direct manipulation~\cite{horvitz99mixedinitiative}, and Shneiderman argues that AI should augment human abilities while maintaining appropriate human oversight rather than pursuing full autonomy~\cite{shneiderman2022human}. 
Amershi et al.\ propose design guidelines stressing user control and error recovery~\cite{amershi19guidelines}. 
The visualization community has similarly sought a balance between agency and automation~\cite{heer19agencyautomation, wongsuphasawat17voyager2, moritz19tvcg}.

\contributionbox{\textbf{Contribution.}
\techname{} preserves user agency throughout the stepwise design process: users may accept a recommendation, override individual parameters, or delegate multiple steps at once. 
This instantiates HCAI principles~\cite{shneiderman2022human} of human oversight over automated processes.
}

\section{Stepwise Visualization Autocompletion}
\label{sec:03-design}

Visualization design is an inherently iterative process~\cite{Munzner2014, parsons22designprac, shin23perceptualpat}: designers begin with only a partial sense of their goal, progressively refine their choices, 
and eventually find an acceptable solution (\autoref{fig:navigation-process} left).
Ideally, they also collect transferable knowledge about designing good charts in the process.
Yet most existing tools force a trade-off: automated tools sacrifice agency in the name of speed, while manual tools demand expertise that domain experts lack~\cite{Wong2018}.
This leads us to our research question: \textit{How can we support the iterative nature of visualization design for novices, enabling meaningful agency while integrating best design practices and transferring visualization knowledge in the process?} 

\subsection{Design Goals and Requirements}
\label{sec:design-considerations}

We identify two complementary goals that any answer to this question must satisfy simultaneously.
\rev{\textbf{Approachability} means that a novice with no background in data visualization can engage with the system immediately, without learning a specification language or mastering a grammar of graphics.
\textbf{Articulacy} in our context, is how faithfully a chart designer realizes a given \textit{communicative} intent (the target message) in a chart.}
The system enables nuanced, intentional expression, that users can steer toward the chart they have in mind rather than accepting whatever the system produces.

Our proposed method for \textit{stepwise visualization autocompletion} achieves both simultaneously: keeping the process approachable by presenting an ordered list of candidate states, and pursuing articulacy by integrating data-driven recommendations with user choice.
This yields five design requirements:

\begin{itemize}[leftmargin=0.77cm]

    \item[DR1]\textbf{Responsive:} Recommendations must be generated in real time~\cite{shneiderman83directmani, wei23autocomplevis}, since the value of autocompletion collapses if the user must wait.

    \item[DR2]\textbf{Understandable:} The interface must require no prior knowledge of visualization grammars or specification languages.

    \item[DR3]\textbf{Anticipatory:} Users must be given a preview of how a recommendation will affect the visualization, so they can make an informed decision before committing to a choice.
    
    \item[DR4]\textbf{Trackable:} Users must be able to see where they are in the design process, and reflect how each step changed the chart.
    
    \item[DR5]\textbf{Controllable:} Users must be able to intervene at any step, \rev{modify the chart,} delegate multiple steps at once, or override any parameter.
    
\end{itemize}

We note that \textit{agency}, the ability to intervene, override, or direct the process, is not a separate goal but an emergent consequence of articulacy: a system that enables fine-grained expression necessarily gives users meaningful control.

\subsection{Generalizing Autocompletion}
\label{sec:generalization}

Our approach is inspired by text autocompletion systems, which suggest probable and high-quality continuations as a user types a query or phrase~\cite{darragh90reactivekeyboard}.
At its core, autocompletion is a \textit{ranked navigation method} through a combinatorial state space: the user anchors a starting point, and the system returns an ordered set of next states according to some model of quality or likelihood (\autoref{fig:navigation-process} right). 
This characterization is not specific to text: any domain that can be modeled as a space of valid states with ranked transitions is a candidate for autocompletion.

Text autocompletion, however, is a single-step interaction: the user selects one completion and issues the query.
Applying the idea to visualization design means that the output is a \textit{persistent artifact}, not a disposable search term.
Because the designer refines an object rather than issuing a one-shot command, the interaction must support \textit{multiple steps}, each one narrowing the design space toward a satisfying result.
We argue that this is a natural generalization: autocompletion applies to any artifact-producing process where (1) the space of valid states is large and difficult to navigate unaided, (2) transitions between states can be ranked by quality, and (3) user intervention at any step is preserved.

\subsection{Formalization: Navigating the Design Space}
\label{sec:statespace}

\begin{figure}[t]
    \centering
    \includegraphics[width=\linewidth]{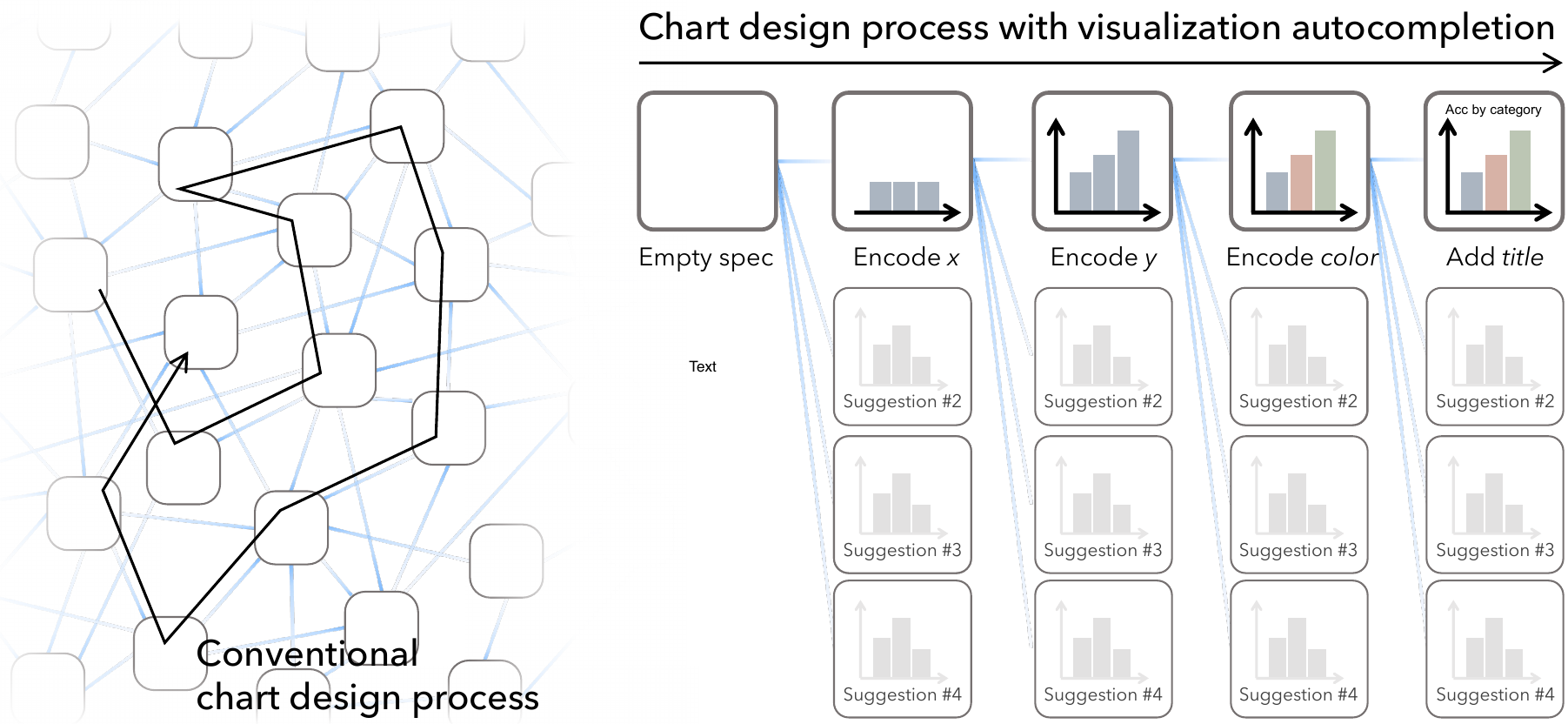}
    \caption{\textbf{Recommendation as design space navigation.}
    Visualization autocompletion redefines visualization recommendation as ranked navigation in the entire space of valid visualization specifications.
    }
    \vspace{-6mm}
    \label{fig:navigation-process}
\end{figure}

To formulate visualization design as a ranked navigation problem (\autoref{fig:navigation-process}), we model the visualization design space as the set of all valid visualization specifications in a given grammar.
Each valid specification $V$ is a \textit{state}, and a \textit{transition} $t$ moves from one valid state to another by making exactly one change.
Formally, each transition $t$ takes a visualization $V$ and user-defined parameters $\theta$ and returns a new specification $V'$, where $V' = t(V, \theta)$.
For example, given a scatterplot $V$ with $x$ and $y$ encodings already defined, a probable transition $t$ is the addition of a color channel, where $\theta$ specifies the choice of encoded variable and color palette.
The resulting structure is a directed graph over a large---potentially infinite---space; 
visualization autocompletion is a method for navigating it.

However, there may be a bewildering number of transitions $t$ available for a given visualization state $V$. 
Visualization autocompletion is thus the problem of recommending, at each state, the highest-quality transitions available.
To make the choice tractable to a novice user, we propose ranking the transitions based on some quality measure.
Ideally, the top-ranked transition may help visualization designers make their charts better reflect the intended message.
This means that the navigation of visualization design space using autocompletion can be viewed as an exploration of a Markov chain in which transition likelihood $P(V' \mid V, m)$ may reflect the quality of $V'$ with respect to a target message $m$.

\section{System: \techname}
\label{sec:04-method}


We present \textsc{\techname}, a visualization authoring system (\autoref{fig:visautocomplete-interface}) that implements our proposed approach of stepwise visualization autocompletion (\autoref{sec:03-design}).
We discuss our recommendation algorithm, an agent mode that automates multiple steps of recommendation, and the design of \techname{} interface.

\subsection{Recommendation Algorithm}

\techname{} relies on an algorithm that recommends a list of transition candidates for each visualization authoring step. 
We describe our recommendation criteria, then detail how we train and implement the criteria as an algorithm.

\subsubsection{Recommendation Criteria}

\label{sec:reccriteria}

Given a visualization specification $V$, \techname{} recommends the list of transitions $t$ ranked by their quality. 
Though ideally the quality of $t$ should be determined by the alignment between $V'=t(V,\theta)$ and the target message $m$, computationally assessing this alignment is nontrivial.
We may leverage LLM's capability to understand chart semantics \cite{ko24chi, tian25chartgpt}, but this approach compromises execution time. Furthermore, designers often lack an explicit ``message'' to convey, making it inappropriate to require such input to the system.

We instead assess the quality of a transition $t$ based on the \textit{plausibility} of the transition, i.e., \textbf{the extent to which average visualization designers would commonly select it}.
Our rationale is that, regardless of the target message $m$, such recommendations can guide authors toward visualizations that are likely to communicate effectively.
Users can then evaluate whether these plausible recommendations align with their implicitly held message and adjust $\theta$ to improve this alignment.

\subsubsection{Implementation}

We detail how we implement our recommendation algorithm.

\paragraph{Visualization grammar}
We use Vega-Lite \cite{satyanarayan17vegalite} for two reasons. 
First, it provides a large, semantically rich design space for authoring visualizations \cite{satyanarayan17vegalite, zong23tvcg, wongsuphasawat16tvcg}, which enables \techname{} to reflect the needs of diverse users and domains. 
Second, the specification structure consists of bricks of high-level \textit{semantic primitives} \cite{satyanarayan17vegalite, VanderPlas18joss} (e.g., mark, encoding, style). 
This is not only easy to read \cite{narechania21tvcg} but also aligns well with our objective to predict `plausible transition' of a given visualization spec \cite{zong23tvcg, kim17chi}.
Still, we argue that the concept of visualization autocompletion is dataset-agnostic: it can be done with other codes or libraries, such as Matplotlib, or Plotly.js.

\paragraph{Base corpus}
We utilize a Vega-Lite dataset collected by Ko et al. \cite{ko24chi} to train our model. This dataset consists of 1,981 Vega-Lite specifications that represent real-world visualizations. 
We use this dataset because it is semantically diverse across multiple dimensions, including task complexity, chart type, and the presence of composite views. 
Such diversity makes it well-suited for uncovering how people commonly author visualizations (\autoref{sec:reccriteria}).

\paragraph{\rev{Recommendation engine}}
We build our \rev{recommendation engine} by (1) decomposing visualization specifications into semantic primitives and (2) computing their co-occurrence patterns. Given a visualization $V$ composed of primitives $T_V = \{t_1, t_2, \cdots, t_n\}$, the engine recommends new transitions $t \notin T_V$ that commonly co-occur with those in $T_V$. The detailed steps for training and computing the final recommendation score are as follows.

\paragraphit{(Step 1) Decomposing visualization specifications}
We analyze each Vega-Lite specification in our corpus, decomposing it into semantic primitives---atomic design choices within a visualization specification, such as a mark type (\texttt{\{"mark": "bar"\}}), a field encoding (\texttt{\{"field": "year", "type": "temporal"\}}), or a color scale (\texttt{\{"scheme": "blues"\}}). 
\rev{We iterate through two stages (details can be found in Appendix C)}: 
\begin{enumerate}[topsep=0pt, itemsep=0pt, parsep=0pt, leftmargin=0pt, itemindent=38pt]

\item[\textbf{Stage 1:}] Decompose the code of each chart specification into semantic primitives while ensuring that the resulting primitives can be categorized according to Wilkinson's Grammar of Graphics taxonomy \cite{wilkinson2011grammar}: \textit{data, transform, scale, coordinate, element,} and \textit{guide}. 
We design the deterministic rule for decomposing chart specifications also based on Vega-Lite documentation.
\item[\textbf{Stage 2:}] Merge primitives that share the same semantics across charts as a taxonomy (e.g., all instances that change the width of the charts (e.g., \texttt{continuousWidth} and \texttt{width}) are merged into a single primitive group). As with Stage 1, we design a rule for merging the primitives.
\end{enumerate}
We iterate stages 1 and 2 by re-categorizing the semantic primitives in the chart corpus using the merged primitive set, and then merging semantically equivalent primitives again.
We repeat the loop until the taxonomy of semantic primitives no longer undergoes changes through three iterations of the loop.
\rev{Note that we do not use LLMs to directly generate primitives to reduce uncertainty and ensure consistency of the extraction. We rather use it to validate and refine the set of primitives extracted by our rules for chart decomposition and merging. When a problem is detected, we update the rules. }
This iterative loop can thus be interpreted as a variant of chain-of-thought prompting~\cite{wei22neurps}. 
To further ensure consistency, we guide LLMs to refer to official Vega-Lite documentation using the model context protocol~\cite{hou26tsem}.
We use OpenAI GPT-5.2 to execute LLMs.

\paragraphit{(Step 2) Computing co-occurrence}
Given our corpus $\mathcal{V}$, where $|\mathcal{V}|$ denotes its cardinality (i.e., the total number of specifications in the corpus), we define the co-occurrence ratio $c(t_1, t_2)$ between primitives $t_1$ and $t_2$ as:

\begin{equation}
    c(t_1, t_2) := \frac{|\{v \in \mathcal{V} \mid t_1 \in v \wedge t_2 \in v\}|}{|\mathcal{V}|}
\end{equation}
We precompute the co-occurrence ratio of all pairs of primitives to reduce the runtime when users use the \techname{} system.

\paragraphit{(Step 3) Computing quality scores for recommendation}
Given the current visualization \(V\) created by the designer through \techname{} and its constituent primitive set \(V_T\), The score of \(t\) is defined as its arithmetic mean of the co-occurrences with the primitives already present in \(V_T\).
Formally, it is represented as: $q(V,t) := \mathrm{mean} \{\, c(t, t') \mid t' \in V_T \,\}$.
Note that before any visualization is specified, the first recommendation is generated immediately after the user uploads a dataset. At this stage, \(V_T\) contains only the primitive representing the dataset type (e.g., tabular, time-series, or geospatial; classified as \textit{data} in Wilkinson's taxonomy), and recommendations are made based on the initial choice made by users.

\subsection{The Agent Mode}

\label{sec:agents}

\techname{} provides \textit{the agent mode}, a multi-step autocompletion to reduce cognitive load and increase efficiency.
We define \textit{agents} as an automated process that proactively explores several steps ahead of users and recommends several resulting authoring paths.
We develop agents with two objectives:
(1) different agents should provide diverse recommendations, both in terms of the types of design choices made and the number of steps taken. 
(2) while doing so, agents should follow common visualization authoring practices in their recommendations.
To this end, we control agent behavior via two hyperparameters ranging from 0 to 1: \textit{randomness} \(r\) and \textit{saturation threshold} \(u\). 

The \textit{randomness} parameter controls the diversity in design choices.
Formally, we define the probability of selecting recommendation \rev{$P_i(r)$} as a softened softmax:
\[
P_i(r) := \frac{e^{s_i(1-r)}}{\sum_{j \in \{1,\ldots,n\}} e^{s_j(1-r)}},
\]
where \(s_1, \ldots, s_n\) denotes the quality scores of top-\(n\) recommendations.
As \(r\) increases, the distribution becomes flatter, allowing the agent to explore lower-ranked recommendations more often. 
We randomly pick \(r\) ranging from 0 to 0.4 to prevent agents from outputting recommendations that do not follow the common practices.

The \textit{saturation threshold} \(u\) determines when the iterative process should saturate and terminate. 
At each iteration, we quantify the difficulty of choosing among current recommendations using the entropy of the zero-randomness distribution \(P_i(0)\):
\[
H = -\sum_{i \in \{1,\ldots,n\}} P_i(0)\log P_i(0).
\]
A higher entropy indicates that the recommendation scores are relatively uniform, suggesting that no transition is clearly preferred, whereas a lower entropy indicates that the choice is more concentrated on a few strong candidates. 
The iteration terminates when \(H < u\). 
We randomly pick \(u\) within the range of 0.2 to 0.5.
The range is determined heuristically. 

When the agent mode is enabled, \techname{} instantiates multiple agents and generates recommendations. 
Users can accept recommendations as is, apply them with minor adjustments such as modifying parameters (e.g., color palettes) or discarding specific steps, or dismiss them entirely.

\subsection{Translating Recommendations into Vega-Lite Code}

One technical challenge in \techname{} is the translation of recommended transitions to Vega-Lite code. 
This is nontrivial because incorporating a new primitive into a Vega-Lite specification often requires more than simply appending a new leaf node; it also requires synchronizing with other parts of the specification. 
For example, when an \(x\)-axis is binned, adding a new variable to the \(y\)-axis requires determining how data points within each bin should be aggregated. 
Similarly, when multiple marks are combined through layering, the encoding of a new mark (e.g., combining \texttt{"confidence interval"} mark with \texttt{"line"} mark) must be determined in a manner that is consistent with the encodings of existing marks.
One trivial way to implement this translation is to call an LLM, but this increases system latency (DR1).
 
To that end, we build a \textit{LLM-distilled translation function} to address this problem. 
Our approach is to prompt the LLMs (OpenAI GPT-5.2) to autonomously design charts using \techname{}, while the LLMs translate each recommended transition into an executable Vega-Lite code. 
For each translation, we store both the resulting code and the contextual information explaining why that translation was produced. 
We then distill these accumulated translations into a single JavaScript function that replicates the LLM's translation logic deterministically.
The function receives a Vega-Lite spec $V$, a recommended transition $t$, and a parameter $\theta$, then outputs a new Vega-Lite spec $V'$ with the transition applied, i.e., $V'=t(V,\theta)$.
By repeatedly running this process across diverse datasets, the function accumulates coverage over a wide range of design sequences, enabling \techname{} to perform code translation responsively without on-the-fly LLM inference.

\begin{figure*}[tbh]
    \centering
    \vspace{-0.5cm}
    \scalebox{0.99}{
    \includegraphics[width=\linewidth]{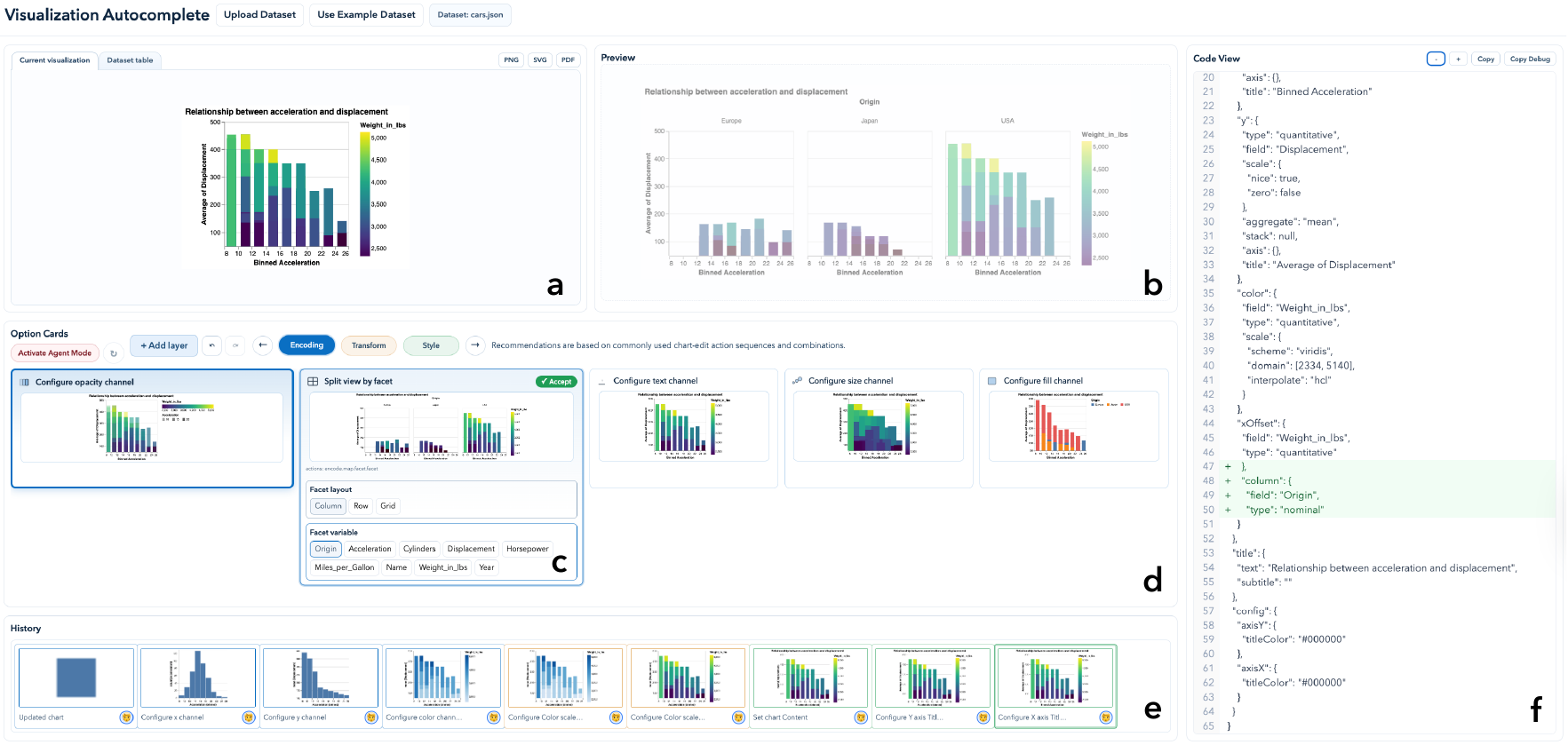}
    }
    \caption{\textbf{The \techname{} interface.} Designers can update the \textit{current visualization} (a) by exploring design recommendations presented in the \textit{recommendation panel} (d). They can further tailor the visualization to their intent by adjusting the parameters of the \textit{recommendation cards} (c). As they explore these options, the system \textit{previews} the resulting visualization design (b) and shows how the visualization specification would change in the \textit{code view} (f). Finally, designers can track how the visualization design evolves through the \textit{graphical history view} (e).
    }
    \vspace{-5mm} 
    \label{fig:visautocomplete-interface}
\end{figure*}

\subsection{The \techname{} Interface}
\label{sec:04-ui}

We describe the \techname{} interface (\autoref{fig:visautocomplete-interface}). 

\paragraph{Interactive spreadsheet}
When users first upload their data, the system automatically shows it in an interactive spreadsheet and displays it in the upper-left panel of the interface. 
Users can then interactively explore the data and select the columns they wish to include in the visualization. 
After making their selections, they proceed to the authoring phase by clicking the `Apply' button. 
The spreadsheet view is then automatically replaced with the chart view.

\paragraph{Recommendation panel}
After users \rev{finish selecting the data} columns of interest, the recommendation panel (\autoref{fig:visautocomplete-interface}d) then presents  top-$K$ (in our case, $K$ was 5, and can be extended) candidates for the next transition as interactive \emph{recommendation cards} (\autoref{fig:visautocomplete-interface}c).
Initially, when the dataset is uploaded, the recommendation shows options to choose the marks.
Afterward, it shows options related to data encoding (e.g., configure $x$ encoding), data transformation (e.g., add \textit{regression} transform), and style (e.g., configure chart \textit{title}). 
Each card includes a thumbnail preview of the visualization produced by accepting the corresponding primitive, along with a short text label that summarizes the primitive's effect.
Cards are ordered by descending score, with the highest quality score appearing at the top-left. 
To help users identify the most commonly chosen options at a glance, the top two cards are visually emphasized with bold borders and increased widths.

Accepting a recommendation is not an all-or-nothing action.
Within each card, below the thumbnail, \techname{} provides direct manipulation controls that users can use to adjust the parameters of the recommended primitive according to their favor (e.g., a recommendation to bin a quantitative field exposes a slider for the number of bins; a color-scale recommendation exposes a palette selector; a title recommendation exposes a text field).
These controls let users accept a recommendation while still adjusting its details to their intent (DR5).
In other words, the default values reflect best-practice guidelines, but can be overridden at any time.
The controls appear only after a card is selected, keeping the panel uncluttered during browsing.

Note that the system first recommends chart marks, then allows users to request recommendations from one of three distinct categories: \textit{encoding}, \textit{data transformation}, and \textit{style}. 
Encoding recommendations specify how attributes are mapped to visual channels, data transformation recommendations apply operations that modify the data directly, such as filtering or scaling, and style recommendations adjust aesthetic properties. 
We adopt this categorization to prevent users from being overwhelmed by recommendations with substantially different characteristics. 
This design also aligns with the stages described in the reference model of Card, Mackinlay, and Shneiderman \cite{card1999readings}.

\paragraph{Chart view and preview}
The right side of the upper panel displays the current Vega-Lite visualization authored by users (\autoref{fig:visautocomplete-interface}a), along with buttons to download it in various formats, including .png and .pdf. On the left side, the ``preview'' of the chart that will be confirmed when the recommendation card is clicked is displayed with lower opacity (\autoref{fig:visautocomplete-interface}b). 
This view is updated instantly as the user hovers over recommendation cards or adjusts parameter controls (DR1).
This anticipatory preview (DR3) allows users to evaluate candidate transitions without committing to them or having visualization knowledge (DR2). 
This capability reduces the cost of exploration.


\paragraph{Graphical history view}
At the bottom of the interface, a horizontal, graphical history interface~\cite{heer08graphicalhistories} (\autoref{fig:visautocomplete-interface}e) provides a visual record of every design decision the user has made (DR4).
Each entry is a thumbnail of the chart as it appeared after a given design step, ordered chronologically from left to right.
The most recent state is anchored at the right edge of the visible area, so that the most recent states are visible without scrolling.
Earlier states overflow to the left and are accessible by scrolling.

Clicking any thumbnail restores the system to the corresponding historical state: the recommendation panel repopulates with the candidates available at that step, the live preview reverts to that chart, and the parameter controls reflect the choices made up to that point.
If the user then accepts a different recommendation from this restored state, the new choice is appended immediately to the right of the selected thumbnail, and all entries to its right that represent the previously explored path are discarded (DR5).
This implements a \emph{tree-structured} version history with a single active branch, making non-linear exploration tractable for novice users.

\paragraph{Code view}
To help users track how each decision maps to the underlying specification (DR4), \techname{} displays the Vega-Lite code corresponding to the current chart state (\autoref{fig:visautocomplete-interface}f).
The code view updates synchronously with the live preview: hovering over a recommendation card changes both the chart and the code simultaneously.
Additions to the specification are highlighted in blue, and removals in red, making it immediately apparent how a candidate decision would modify the current state.
\rev{Note that this view is optional rather than core to the novice workflow. 
It serves as a stepping stone for users who wish to learn Vega-Lite, and helps them locate where an error occurred when the tool produces an unexpected result.}

\section{Evaluation}
\label{sec:05-experiment}

We evaluate \techname{} through a controlled user study. Below we describe the evaluation protocol.

\subsection{Objectives}
\label{sec:05-objectives}

We evaluate the effectiveness of \techname{} along two dimensions that reflect the core goals of our system (\autoref{sec:03-design}), by comparing it against widely used visualization authoring systems and automated chart recommendation tools. 
First, we verify that \techname{} is at least as \textbf{approachable} as existing tools, given that its stepwise interface is designed to lower the barrier to entry for workers with low visualization literacy.
\rev{Second, beyond ease of use,  we ask whether \techname{}'s stepwise approach actually improves users' \textbf{articulacy}, i.e., helps users produce visualizations that best delivers users' communicative intent.}
We establish three hypotheses:
\begin{itemize}[leftmargin=16pt]
    \item[\textbf{H1}] Charts produced with \techname{} are of higher quality than those generated by baselines.
    \item[\textbf{H2}] The effect in H1 is amplified when authoring more complex charts (e.g., charts encoding a larger number of variables).
    \item[\textbf{H3}] The approachability of \techname{} is comparable to that of baselines.
\end{itemize}
Here, we define charts as having higher quality if they more effectively convey their intended message. 
We also say a chart authoring tool is more approachable when users feel less cognitive load.

\subsection{Study Design}

To verify these hypotheses, we set two independent variables:
\begin{itemize}[leftmargin=16pt]
    \item \textbf{Visualization authoring tools}: \textit{Visualization Autocomplete} and baseline techniques (\textit{Excel}, \textit{LLM vibecoding}, and \textit{TaskVis} \cite{shen21eurovis})
    \item \textbf{Complexity of charts}: \textit{Simple} (visualizing two features in a dataset) and \textit{Complex} (visualizing four features),
\end{itemize}
examining how these variables impact the quality of charts and approachability. The detailed study design is as follows:

\paragraph{Participants}
Our target population is domain experts who are proficient with data but have little background in visualization design~\cite{Wong2018}---a group that constitutes a significant proportion of everyday chart creators~\cite{dvs2024}.
We recruit 18 office workers (P1--P18) with experience creating charts for communicative purposes in their daily workflow, but without formal training in data visualization (10 males and eight females, aged 23--42 [$30.5 \pm 6.5$]). The average experience of authoring visualizations ranges from 1 to 10 years ([$4.3 \pm 2.8$]). 
We leverage the internal web community of three local universities and snowball sampling \cite{goodman61ams}.  We constrain participants to have experience in creating charts for communicative purposes in their daily workflow. We compensate the participants with a gift card equivalent to 15 USD.

\paragraph{Assignment}
For each trial, we provide participants with (1) a dataset, (2) a message, and (3) a visualization authoring tool.
Then, we ask them to create a chart that effectively communicates the message to a general audience.
We preprocess the dataset and remove features unrelated to the target message to minimize the influence of differences in data literacy on the study outcomes. 
We develop a website that allows participants to submit the final version of their visualization by uploading the relevant files or copy-pasting a screenshot.

\paragraph{Baseline techniques}
We evaluate the superiority of \techname{} relative to existing authoring tools that provide machine support. 
We select baselines that span a range of agency levels and approachability. 

\paragraphit{Microsoft Excel}
We include Microsoft Excel as a baseline because it is the most widely used tool for data analysis and chart creation among office workers \cite{chal22chi}.
It is also a representative tool with relatively low user agency and high approachability. When using Excel, we provide data in both long and wide formats, and instruct basic functionalities like sort, filter, and transpose. We present participants with the ``Insert'' tab, which includes the ``Recommended Charts'' feature, and encourage them to use this functionality.

\paragraphit{LLM vibecoding}
\rev{We include LLM vibecoding (e.g., ChatGPT or Claude) as a baseline and instruct participants to use natural language prompts to generate visualizations. We add this baseline because prompt-based visualization authoring has become increasingly common \cite{shen23tvcg, wu24pamd}.
We want our baseline to best represent what office workers without visualization expertise would most commonly use.
We select ChatGPT because it is currently one of the most widely used commercial LLM services \cite{russell25acl, chen25msom}
\footnote{\href{https://www.techbusinessnews.com.au/news/92-of-fortune-500-companies-use-openai-products/}{\texttt{https://www.techbusinessnews.com.au/...}}}
\footnote{\href{https://www.reuters.com/technology/openais-altman-pitches-chatgpt-enterprise-large-firms-including-some-microsoft-2024-04-12/}{\texttt{https://www.reuters.com/...}}}
. We also consider programming assistants like Codex or Claude Code, but exclude them because they are less commonly used by office workers and more likely to target expert programmers.
We use the official service interface\footnote{\url{https://chatgpt.com/}} with a ChatGPT Pro subscription.
}
We use ``developer mode'' to disable memory functionality, constraining the LLM not to reference previous chat logs. 
We do not constrain the LLM's behavior, e.g., by using a system prompt, to simulate regular office work. 
\rev{We conduct the experiment from March 21st to 26th, 2026, where participants can freely select ``Instant'' or ``Thinking'' mode with GPT-5.3.}

\paragraphit{TaskVis}
We include TaskVis \cite{shen21eurovis} as a representative for an automated visualization recommendation engine. We selected TaskVis over alternative systems such as Draco \cite{moritz19tvcg} and Dziban \cite{lin20dziban}.
This is because it allows the analytical tasks that a visualization should support (e.g., ``correlate'' or ``compare'') to be specified explicitly as an input parameter, which aligns well with our study design, where participants create charts based on a given message.
Because TaskVis is a fully automated system and also requires users to understand its syntax, we do not ask participants to use it directly. Instead, we execute it ourselves by translating the prepared chart messages into corresponding tasks and passing them to the TaskVis module as input parameters.

\paragraph{Stimuli design}
We design our stimuli (chart messages and corresponding datasets) to simulate everyday analytic contexts, or situations in which an office worker creates charts to communicate messages to the general public or to coworkers with limited data literacy. 
To this end, we derive stimuli by examining visualizations published in news articles.
The detailed procedure is as follows:

\paragraphit{(Step 1) Investigating reference dataset}
We select news visualizations from the MASSVIS dataset~\cite{borkin13tvcg} as our reference corpus. 
To minimize bias and reduce manual effort, we extract 200 semantically diverse charts via stratified sampling \cite{pandey16chi, abbas19cgf} (see Appendix A for details) from the corpus. 
We then manually analyze these charts, extracting (1) the domain, (2) the number of variables, (3) the intended message, and (4) the analytical tasks required to decode the message. 
Two coders independently analyze the charts and reconcile their annotations through two discussion sessions. 
Initial inter-coder agreement, measured by Cohen’s $\kappa$, is 0.58. 
Note that while classifying analytical tasks, we refer to Quadri and Rosen's taxonomy on visualization tasks \cite{qaudri22tvcg}.

\paragraphit{(Step 2) Selecting analytical tasks}
To determine the analytical tasks that participants' charts should support, we analyzed the frequency of tasks in our reference dataset. 
As a result, we select \texttt{characterize distribution}, \texttt{sort}, and \texttt{correlate} as representative analytical tasks, as they are (1) the most commonly appearing tasks in the corpus, and (2) frequently used in isolation without being paired with other tasks. 
Although \texttt{compare} is the most frequent task in our corpus, it is inherently combinatorial; it rarely appears in isolation but instead augments other tasks (e.g., comparing distributions, comparing correlations). 
We therefore exclude it from simple charts, where each trial involves a single analytical task, and incorporate it into complex charts, where multiple tasks are combined.
Hence, each participant completed all combinations of analytical tasks and independent variables, resulting in 3 [analytical tasks] $\times$ 3 [authoring tools] $\times$ 2 [complexity levels] = 18 trials per participant.

\paragraphit{(Step 3) Selecting datasets and chart messages}
The intent behind selecting datasets and messages is to ensure semantic diversity across chart messages, to minimize the influence of participants' domain background knowledge on the experimental results.
We adopt nine major news topics from The New York Times\footnote{\url{https://www.nytimes.com/international/}} (Domestic, World, Business, Lifestyle, Art, Sports, Games, Cooking, and Economy). 
For each analytical task (\texttt{characterize distribution}, \texttt{sort}, and \texttt{correlate}), we randomly assign three non-overlapping topics.

For each topic-task pair, we then collect datasets suitable for creating visualizations that support the assigned analytical task.
To do so, we first manually collect 30 datasets per topic from public repositories such as Kaggle. We then filter out datasets that are not suitable for the assigned task (Appendix A for detailed criteria).
For example, for \texttt{correlate}, we retain only datasets containing at least two quantitative variables with a Pearson correlation greater than 0.5. 
We then examine the remaining datasets and curate candidate chart messages that can be derived from each dataset. 
From these, we select one message and its corresponding dataset for a simple chart and another for a complex chart. 
When multiple candidate messages are available, we prioritize messages that can be expressed through a diverse range of chart designs.
Messages for simple and complex charts involve two and four variables, respectively. 
These correspond to the 25th and 75th percentiles of the variable count distribution in our reference corpus. 
The former reflects commonly created charts and the latter represents more demanding charts that office workers may occasionally encounter.

\paragraph{Procedure}
After participants sign the consent form, we explain the study purpose and tasks. 
Participants then complete three sessions; each session comprises six trials in which they author charts using one of the three tools (\techname{}, Excel, and an LLM). 
We counterbalance the tool order across participants using a full-factorial design.
After each session, participants are guided to evaluate the approachability of the tools using NASA-TLX and the system usability scale (SUS).
Within each session, three trials require participants to create complex charts and the other three require simple charts. 
We fully randomize trial order to counterbalance complexity. 
We counterbalance the assignment of chart messages across sessions using a Latin-square design. 
We conduct a post-hoc interview after all sessions are finished. 
All studies have finished within 90 minutes. 

\paragraph{Evaluating articulacy}
We evaluate chart quality as a proxy for articulacy. 
To this end, we recruit eight visualization experts to assess the charts created with each tool (all male, aged 27--45, $35.6 \pm 6.4$). 
All evaluators hold a Ph.D. degree, including three professors, four postdoctoral researchers, and one industry researcher. 
Their primary research area is data visualization, where they regularly publish in major visualization venues such as TVCG and VIS. 
On average, they have 10.1 years of experience in visualization research ($\pm 3.8$).

After they sign the consent form, we provide evaluators with the charts made by study participants and the corresponding message. 
\rev{Aligned with our definition of articulacy, }we ask them to determine the articulacy based on the degree to which the chart effectively and accurately conveys the message to general audiences.
We show each pair of charts and messages individually, and guide evaluators to answer the Likert scale questionnaire spanning from 1 (very poor) to 7 (very effective). 
All evaluations have finished within 40 minutes. 
We compensate each participant with a USD 20 gift card. 

We sampled 40 charts per tool (160 total) using stratified sampling to balance topic and complexity, with each chart rated by three experts, yielding 480 evaluations (120 per tool).


\section{Results}
\label{sec:06-study}

We present our quantitative evaluation results on articulacy and approachability. 
We then discuss the usage patterns of \techname{}, as well as the qualitative findings from the post-hoc interview and charts created by study participants.

\subsection{Quantitative Results}

\label{sec:quant}

We discuss our quantitative findings on articulacy and approachability in using \techname{} and baselines.

\begin{figure}
    \centering
    \includegraphics[width=\linewidth]{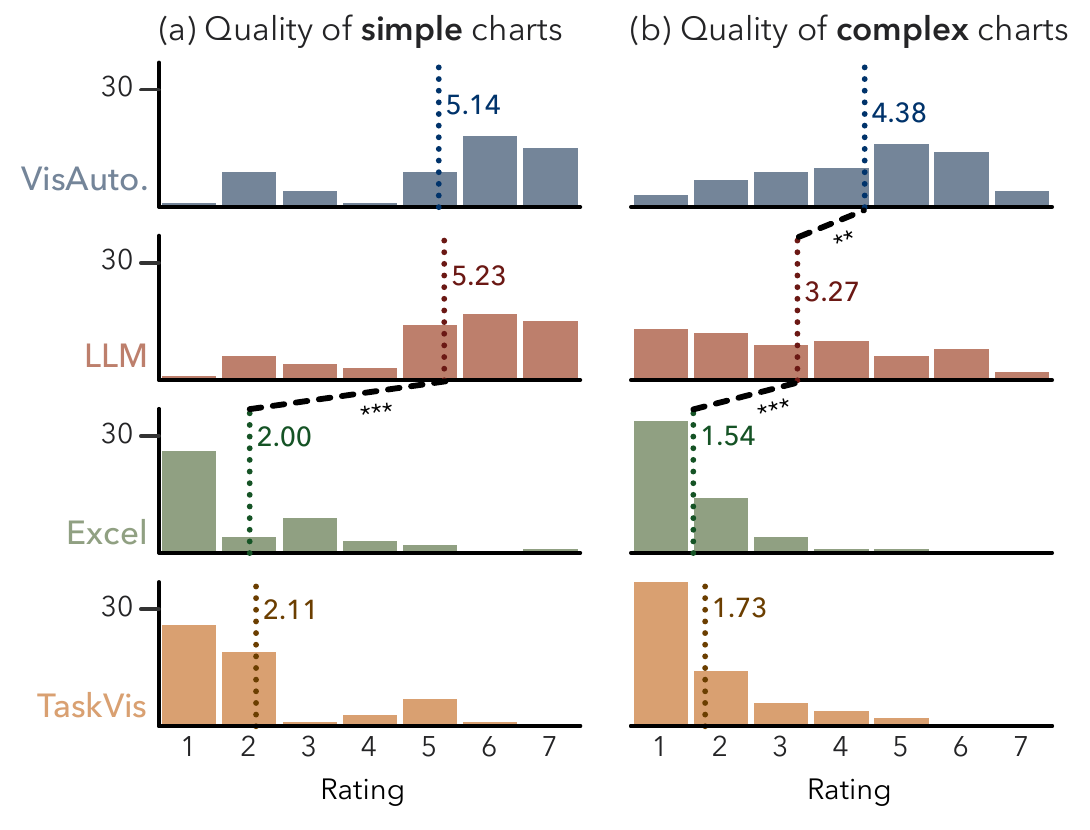}
    \caption{\textbf{Articulacy of charts created by participants.}
    \techname{} produced higher-articulacy charts than Excel and TaskVis, and outperformed LLM vibecoding for complex charts (***: $p < .001$, **: $p < .01$).}
    \label{fig:articulacy}
    \vspace{-7mm}
\end{figure}


\paragraph{Articulacy}
We conduct a two-way Aligned Rank Transform (ART) ANOVA to examine the effects of the independent variables, visualization authoring tool and chart complexity, on chart quality. 
We use ART ANOVA because chart quality is assessed with Likert-scale questionnaires, making a nonparametric analysis appropriate. 
We also use ART contrast tests with Holm correction for post-hoc analysis.

As a result (\autoref{fig:articulacy}), \techname{} consistently produces higher-quality charts than TaskVis and Excel, and further outperforms LLM vibecoding when authoring complex charts (confirms H1, H2).

We find that the selection of visualization authoring tools significantly alters the chart quality ($F_{3,143.73} = 77.62$, $p < .001$).
Post hoc analysis shows that both \techname{} and the LLM vibecoding produced significantly higher-quality charts than Excel and TaskVis ($p < .001$ for all comparisons). 
We also find that chart complexity significantly affects chart quality ($F_{1,145.93} = 42.03$, $p < .001$), with complex charts receiving lower scores than simple ones ($p < .001$).
Furthermore, the interaction between authoring tool and chart complexity is also statistically significant ($F_{3,144.16} = 5.59$, $p < .01$).
To further examine this interaction, we conduct separate one-way ART ANOVAs on visualization authoring tool for simple and complex charts. 
For simple charts, the ANOVA and post hoc results are consistent with those for the full dataset. 
For complex charts, however, \techname{} outperforms not only Excel and TaskVis ($p < .001$ for both), but also the LLM vibecoding ($p < .01$) in chart quality.




\begin{figure*}[ht]
    \centering
    \includegraphics[width=\linewidth]{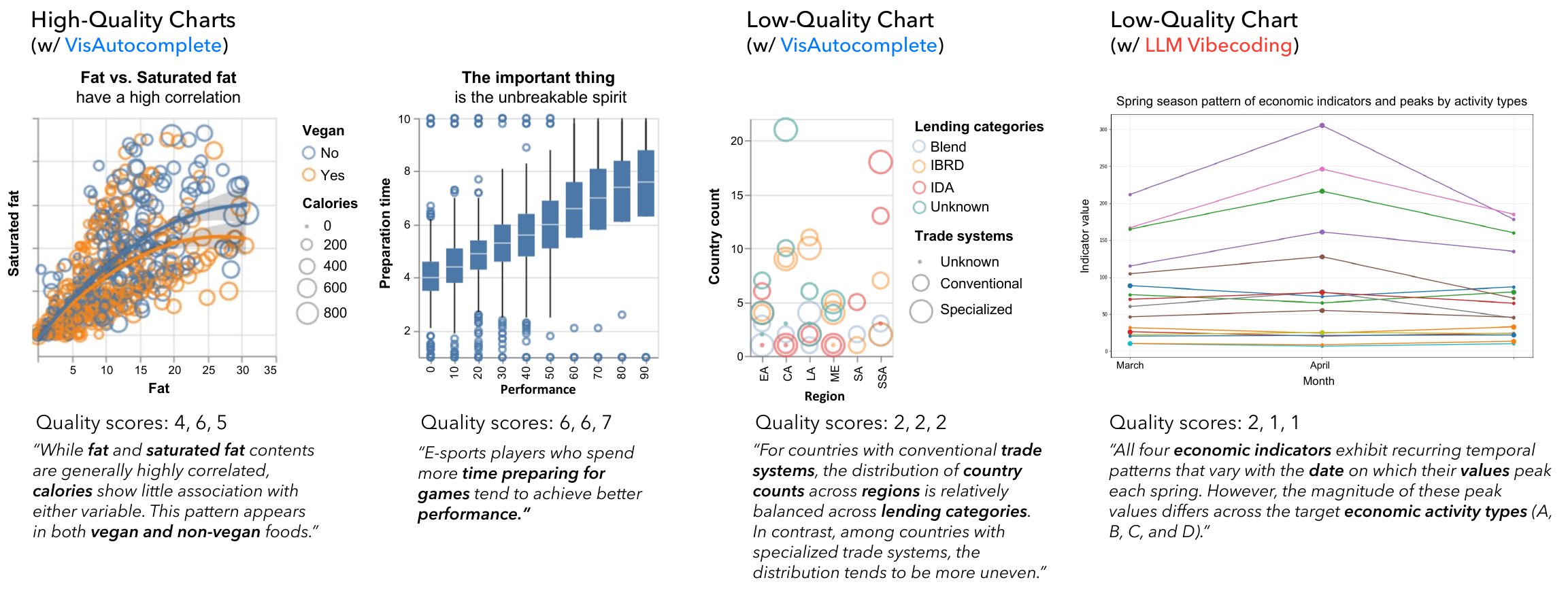}
    \caption{\textbf{Example charts designed by participants in our user study.}
    The quality scores of each chart evaluated by visualization experts are shown below the chart, followed by its intended message in italics. Note that the study is conducted in Korean, and the chart text shown in the figure has been translated into English from the original outputs.}
    \vspace{-3mm}
    \label{fig:example}
\end{figure*}

\begin{figure*}[ht]
    
    \centering
    \includegraphics[width=\linewidth]{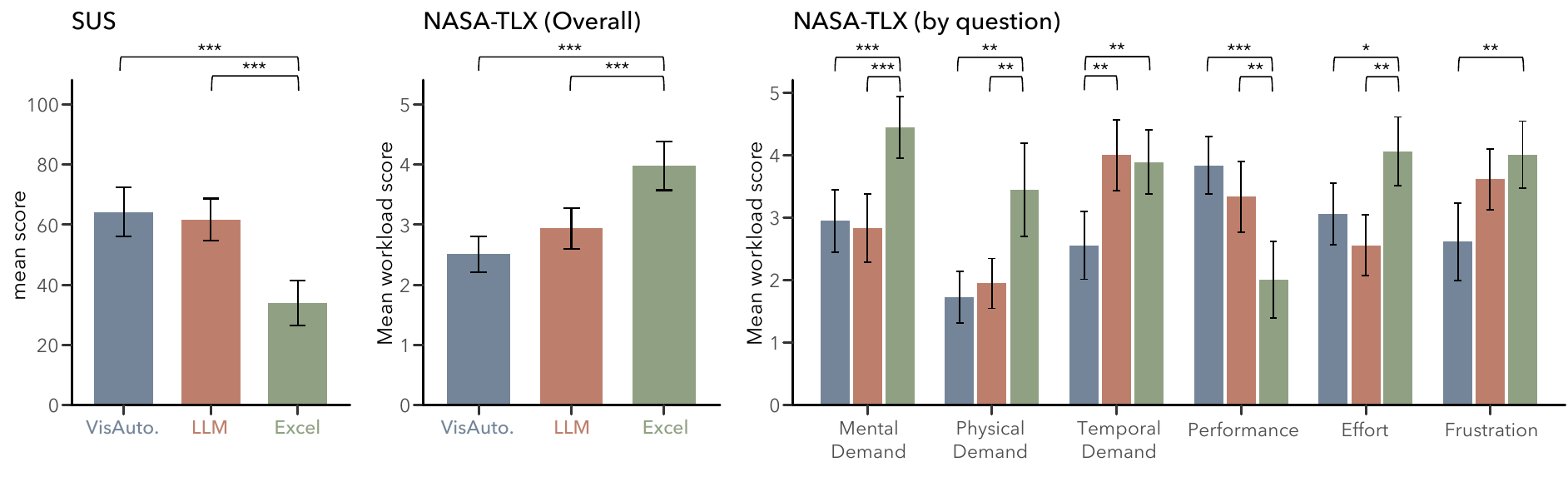}
    \caption{\textbf{Usability of visualization authoring tools measured using SUS and NASA-TLX.}
    Although \techname{} and the LLM vibecoding exhibited similar overall usability, the factors that reduced their usability differ (***: $p < .001$, **: $p < .01$, *: $p < .05$). }
    \label{fig:usability}
    \vspace{-5mm}
\end{figure*}

\paragraph{Approachability as usability scores}
We conduct one-way ART ANOVAs to examine (1) SUS scores, (2) average NASA-TLX scores, and (3) scores on individual NASA-TLX items. For consistency in interpretation, we reverse the \texttt{performance} item, as lower values originally indicate better performance.

As a result (\autoref{fig:usability}), \techname{} and the LLM vibecoding achieve comparable usability, both outperforming Excel, though the LLM imposes greater temporal demand than \techname{} (confirms H3).
The selection of authoring tools significantly changes both SUS scores ($F_{2,34}=15.93$, $p < .001$) and average NASA-TLX scores ($F_{2,34}=19.30$, $p < .001$).
Post hoc analysis shows a similar pattern for both measures: \techname{} and the LLM vibecoding exhibit comparable usability, and both outperform Excel ($p < .001$ for all pairwise comparisons). This overall pattern is also largely consistent across the individual NASA-TLX items.
However, the factors underlying reduced usability differ between \techname{} and the LLM vibecoding. In particular, the LLM vibecoding imposes significantly higher temporal demand compared to \techname{} ($p < .01$).



\paragraph{Task completion time}
We conduct a one-way repeated measure (RM) ANOVA to examine differences in task completion time across visualization authoring tools. 
As a result, we find no significant difference in completion time by authoring tools ($F_{2, 321} = 0.368$, $p = 0.692$). 

\subsection{\rev{Usage Patterns}}
\label{subsec:usage-patterns}
\rev{We provide two analysis pertaining to user agency in \techname{}.}

\begin{figure}[t]
    \centering
    \includegraphics[width=\linewidth]{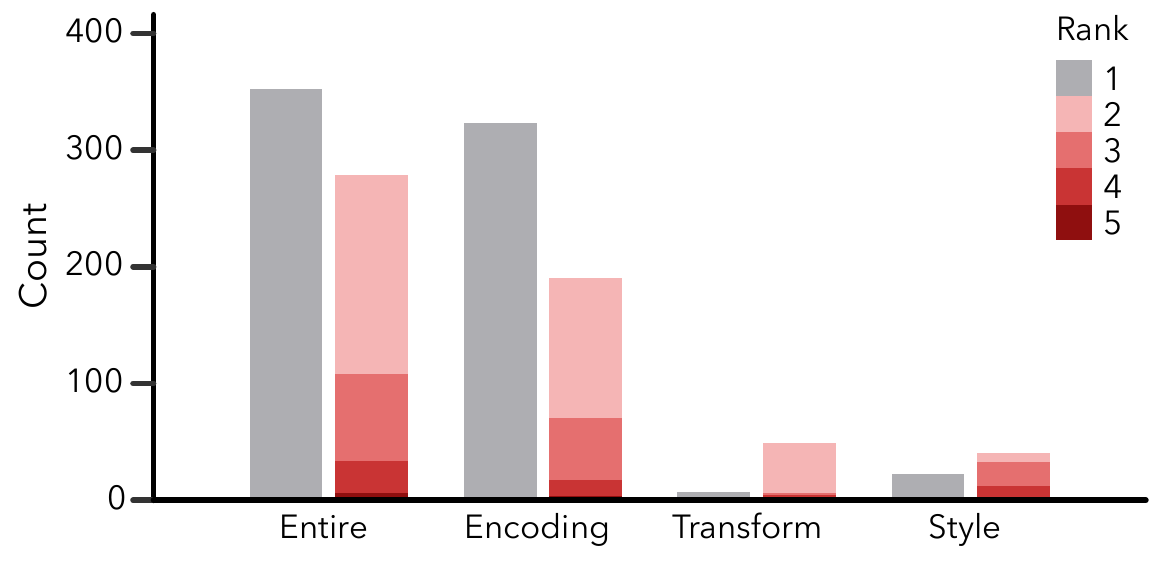}
    \caption{\rev{\textbf{Distribution of the ranks of recommendation cards selected by participants in our user study.} The degree to which participants followed the top-ranked (rank-1) recommendation decreased for data transformation and style. }}
    \label{fig:rec_priority}
    \vspace{-7mm}
\end{figure}

\paragraph{\rev{Reliance on recommendation rank}}
\rev{
To understand how much users rely on the order of recommended charts, we examined which rank they most often selected (see \autoref{sec:04-ui}) across three recommendation categories, including encoding, data transformation, and style.
The results are shown in \autoref{fig:rec_priority}. 
Participants did not simply accept the recommendations presented to them: they declined the top-ranked recommendation in roughly three out of every seven selections. 
Instead, they exercised different degrees of agency depending on the authoring context. 
Because recommendations for data transformation and style often involve more fine-grained design decisions, participants demonstrated greater agency when refining such details, becoming less likely to follow high-ranked recommendations in these categories.}

\rev{
We believe this shift in the dominance of the first recommendation follows from coverage in our dataset. 
Because encoding primitives appear in nearly every chart (e.g., $x$- and $y$-axes), our co-occurrence statistics for encoding are well-estimated, so the top-ranked recommendation more reliably matches what users want. 
Style and data-transformation primitives, in contrast, vary more across charts and thus occur less often individually.
With less data to estimate their co-occurrence patterns, the top-ranked recommendation is more likely to diverge from an individual's actual preference, making participants less likely to follow it.
}

\paragraph{\rev{Restoration behavior}}
\rev{To further examine how actively participants engaged with the tool, we counted the number of restorations, i.e., instances where participants clicked a thumbnail in the graphical history view to revert to a previous state (see \autoref{sec:04-ui}).}

\begin{figure}
    \centering
    \includegraphics[width=\linewidth]{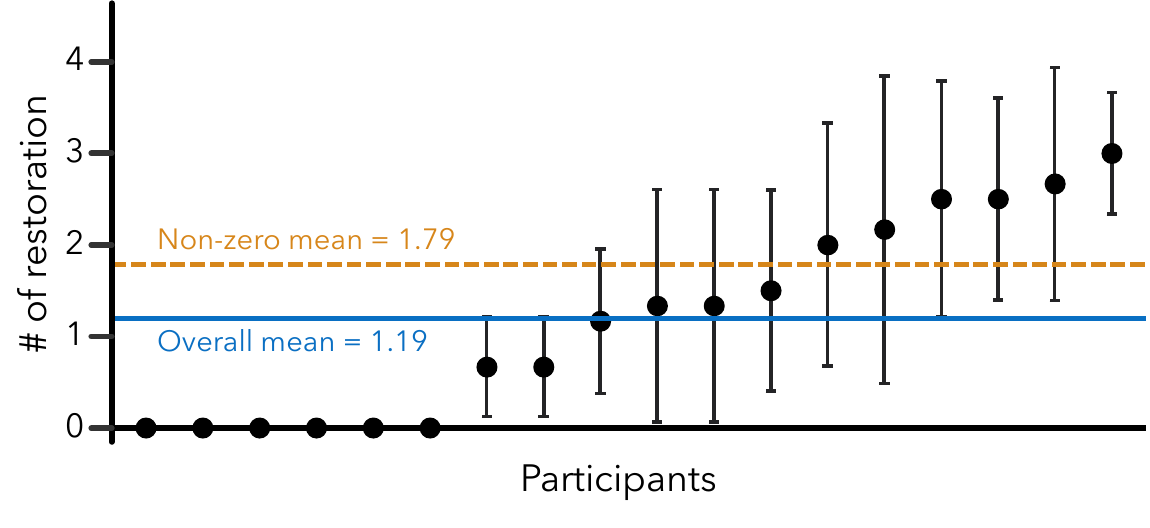}
    \caption{\rev{\textbf{The average number of restoration made in our experiments by participants.} The degree which participants restore the previous version of charts vary substantially by individual. }}
    \label{fig:restore}
    \vspace{-6mm}
\end{figure}

\rev{\autoref{fig:restore} shows the average number of iterations per chart for each participant. 
Restoration use varied substantially across participants: six never restored a previous version, while those who did so restored 1.79 times on average. 
This echoes our analysis of card ranking (above): some participants followed the system's suggested trajectory, while others actively revised it, differing in their desired level of agency. 
\techname{} accommodated both, allowing users to intervene to the extent they wanted.
}


\subsection{Post-study Interview Results}

We discuss notable qualitative results from the post-hoc interview. 

\paragraph{Importance of responsiveness}
Ten of the 18 participants (56\%) note that the long response time of the LLM made them hesitant to iteratively refine the output visualization. 
Among them, three explicitly identified latency as the primary source of frustration when the LLM produced a low-quality visualization. 
In contrast, four participants (22\%) report that \techname{}'s responsiveness motivated them to explore a wider range of design alternatives. 
For example, P3 remark, \textit{``I liked that the visualization changed immediately whenever I hovered over something... That pushes me to explore a variety of options.''}

\paragraph{Needs for explainability}
Participants identified two limitations related to explainability. 
First, four out of the 18 participants (22\%) report that they found it difficult to accept \techname{}'s recommendations because the system did not explain why those recommendations are made. 
Second, six participants (33\%) note that the system provided insufficient explanation of key terms such as mark, encoding, and transform, suggesting that the interface may be difficult to use for users with low visualization literacy. 

\paragraph{Increased efforts with \techname{}}
14 of the 18 participants (77\%) report that they invest the greatest effort when using \techname{} among the three visualization authoring systems, whereas the remaining four pick the LLM vibecoding. Among those participants, five mention that the visual presentation of diverse design options induced them to do so. For example, P11 remarks, \textit{``As I explored various design options using hovering, I suddenly found myself thinking more deeply about which option leads to a good chart design.''} 
This finding is consistent with our quantitative results that charts created with \techname{} achieve higher quality (\autoref{sec:quant}).

\section{Discussion}
\label{sec:07-discussion}

Below we discuss issues on stepwise visualization autocompletion.

\subsection[Articulacy and Approachability]{Approachability and Articulacy} 

On approachability, \techname{} performed comparably to LLM vibecoding and significantly better than Excel, whose limited recommendation scope and lack of previews left users with little support for exploration, causing them to disengage earlier. 
Although \techname{} and the LLM showed no significant difference on SUS or NASA-TLX, interviews revealed qualitatively different experiences.
\techname{} required more effort to use, while LLM vibecoding brought greater time pressure, with participants noting that \textit{``waiting for ChatGPT to generate a chart takes too long.''}

On articulacy, \techname{} and the LLM performed similarly for simple charts, where the design space is small enough that a single well-generated attempt often lands near an acceptable solution. 
For complex charts, however, \techname{} significantly outperformed the LLM and produced fewer lowest-scoring charts, an effect we attribute to the stepwise interface supporting broader exploration at low latency, rather than committing users to a single generated result (see \autoref{fig:example} for failure cases). 
TaskVis underperformed relative to all interactive tools, likely due to its inability to capture task intent without user input (see Appendix B).

\subsection{Chart Authoring as Iterative Search}

Our results and interviews suggest that users approach chart authoring as an iterative search rather than a one-shot specification task. 
Several participants could recognize when Excel's or the LLM's output did not match their intent, even when they could not articulate why or produce a better alternative themselves. 
This suggests that for novices, the challenge may be less about judgment than production. 
Evaluating designs may come more naturally than generating them.

Because users may not be able to directly express their ideal chart, finding it likely requires \textit{exploration}. 
\techname{} supports this by decomposing the design process into sequential steps grounded in common practice, turning authoring into a series of recognition judgments rather than an open-ended generative problem. \rev{Consistent with this, participants declined top-ranked recommendations in roughly three of seven selections and varied widely in restoration behavior (\autoref{subsec:usage-patterns})}, and this exploratory behavior translated directly into higher-quality outcomes for complex charts. 
Giving designers more options to consider, without imposing additional burden, thus helps them find charts that better match their communicative intent.



\subsection{The Impact of Latency in Iterative Design}




A consistent theme across our results is that responsiveness shapes not just user experience, but the iterative design process itself. 
When each iteration is costly, users settle early, accepting a chart that does not fully match their intent rather than exploring further. 
LLM latency discouraged re-iteration after a poor first attempt; as P3 notes, \textit{``I felt deincentivized to iterate with GPT... So I tried to put every possible effort into an initial attempt.''}

A slow tool does not just frustrate users; it curtails the iterations they are willing to make, and therefore how thoroughly they explore the design space. 
\techname{}'s responsive stepwise interface addressed this by making each step fast and lightweight, lowering the cost of exploration until browsing felt natural rather than burdensome. 
In iterative design, latency appears to impact design quality, not just user experience.

\subsection{Stepwise Design Raises the Articulacy Lower Bound}

Beyond producing charts with higher articulacy on average, \techname{} produced notably fewer low-articulacy charts in the complex condition compared to LLM vibecoding. 
We attribute this to a structural property of the stepwise process: comparing several options before committing, rather than receiving a single generated result.

LLM vibecoding, by contrast, delivers a complete chart in a single step. 
If that chart is poor and latency discourages re-iteration, the articulacy of the final chart is bounded by that first attempt. 
As one participant (P11) notes, \textit{``If ChatGPT gets the chart wrong the first time, I don't want to try again. It just takes too long.''} 
Stepwise design raises this lower bound by presenting several recommendations at each step for users to compare. 
\rev{Notably, this persisted even with our suboptimal, coverage-limited scoring (\autoref{subsec:usage-patterns}), suggesting the interaction itself, not scoring precision, is what matters.}

\subsection{Limitations and Future Work}

\rev{We emphasize that our stepwise design recommendation workflow opens up interesting future work directions.
The first direction concerns a new form of design feedback that our decomposition makes possible.
Because the process is decomposed into primitive steps, feedback can target each decision as it is made, rather than only evaluating completed charts as most existing tools do~\cite{shin25visualizationary}.
This matters because preview-based selection alone proved insufficient in our study.
Participants browsed options without developing the confidence to commit.
Step-level feedback is one way to build that confidence, though designing it well raises open challenges.
It must strengthen decision-making rather than just report on it.
It must also keep cognitive load and interruption low.
And it should ultimately advance visualization literacy so that users grow less dependent on it over time.}

\rev{The second direction concerns improving the components in \techname{}.
While \techname{} outperforms existing tools for complex charts, articulacy still declined for complex charts relative to simple ones.
This was particularly true when the large number of variables and encoding choices led novices to suboptimal decisions (see \autoref{fig:example}).
This is consistent with \autoref{subsec:usage-patterns}, where reliance on the top-ranked recommendation dropped precisely for the sparsely covered style and transformation categories.
Addressing these will likely require richer models of design intent that move beyond co-occurrence frequency toward representations capturing a visualization's communicative goals.
Pursuing these directions would bring us closer to enabling users without visualization expertise to author effective charts.}

\acknowledgments{%
    This work was supported partly by Villum Investigator grant VL-54492 by Villum Fonden, and by the Sogang University Research Grant of 202610010.01.
    Hyeon Jeon is in part supported by the Google Ph.D. Fellowship.
    Any opinions, findings, and conclusions expressed in this material are those of the authors and do not necessarily reflect the views of the funding agency.
}

\bibliographystyle{abbrv-doi-hyperref}
\bibliography{vis-autocomplete}

\end{document}